\documentclass[journal=nalefd,manuscript=letter]{achemso}
\setkeys{acs}{keywords = true} 
\usepackage{hyperref,amsmath,mathtools,graphicx}
\usepackage{caption}
\usepackage[T1]{fontenc}
\usepackage{lmodern}
\setkeys{acs}{articletitle=true}
\title{Temperature Dependent Non-linear Damping in Palladium Nano-mechanical Resonators}
\date{\today}

\author{Shelender Kumar}
\affiliation{Department of Physical Sciences, IISER Mohali, Knowledge City, Sector 81, SAS Nagar, Manauli P.O. 140306, India}
\author{S. Rebari}
\affiliation{Department of Physical Sciences, IISER Mohali, Knowledge City, Sector 81, SAS Nagar, Manauli P.O. 140306, India}
\alsoaffiliation{\tiny Present address:Department of Physics, Dr. B.R. Ambedkar National Institute of Technology Jalandhar, Jalandhar 144011, Punjab, India }
\author{Satyendra P. Pal  }

\affiliation{Department of Physical Sciences, IISER Mohali, Knowledge City, Sector 81, SAS Nagar, Manauli P.O. 140306, India}
\alsoaffiliation{\tiny Present address: Nanoscale Research Faciltiy, Indian Institute of Technology,
	Hauz khas, New Delhi 110 016 India. }
\author{S.S. Yadav}
\affiliation{Department of Physical Sciences, IISER Mohali, Knowledge City, Sector 81, SAS Nagar, Manauli P.O. 140306, India}

\author{Abhishek Kumar} 
\affiliation{Department of Physical Sciences, IISER Mohali, Knowledge City, Sector 81, SAS Nagar, Manauli P.O. 140306, India}
\alsoaffiliation{\tiny Present Address : NEST, Istituto Nanoscienze-CNR and Scuola
	Normale Superiore, Piazza San Silvestro 12, 56127 Pisa, Italy.}

\author{A. Aggarwal}

\affiliation{Department of Physical Sciences, IISER Mohali, Knowledge City, Sector 81, SAS Nagar, Manauli P.O. 140306, India}
\alsoaffiliation{\tiny Present address: Department of Physics and Astronomy, Northwestern University, Evanston, Illinois 60208, USA }

\author{S. Indrajeet}

\affiliation{Department of Physical Sciences, IISER Mohali, Knowledge City, Sector 81, SAS Nagar, Manauli P.O. 140306, India}
\alsoaffiliation{\tiny Department of Physics, Syracuse University, Syracuse, NY 13244, USA}

\author{A. Venkatesan}
\affiliation{Department of Physical Sciences, IISER Mohali, Knowledge City, Sector 81, SAS Nagar, Manauli P.O. 140306, India}
\email{v_ananth@rocketmail.com}

\begin{document}
\maketitle
\newpage
\begin{abstract}
Advances in nano-fabrication techniques has made it feasible to observe damping phenomena beyond the linear regime in nano-mechanical systems. In this work, we report cubic non-linear damping in palladium nano-mechanical resonators. Nano-scale palladium beams exposed to a $H_2$ atmosphere become softer and display enhanced Duffing non-linearity as well as non-linear damping at ultra low temperatures. The damping is highest at the lowest temperatures of $\sim 110\: mK$ and decreases when warmed up-to $\sim 1\textrm{ }K$. 
We experimentally demonstrate for the first time a temperature dependent non-linear damping in a nano-mechanical system below 1 K. It is consistent with a predicted two phonon mediated non-linear Akhiezer scenario for ballistic phonons with mean free path comparable to the beam thickness.
This opens up new possibilities to engineer non-linear phenomena at low temperatures. \\

\centering
{
	\scalebox{0.55}{\includegraphics{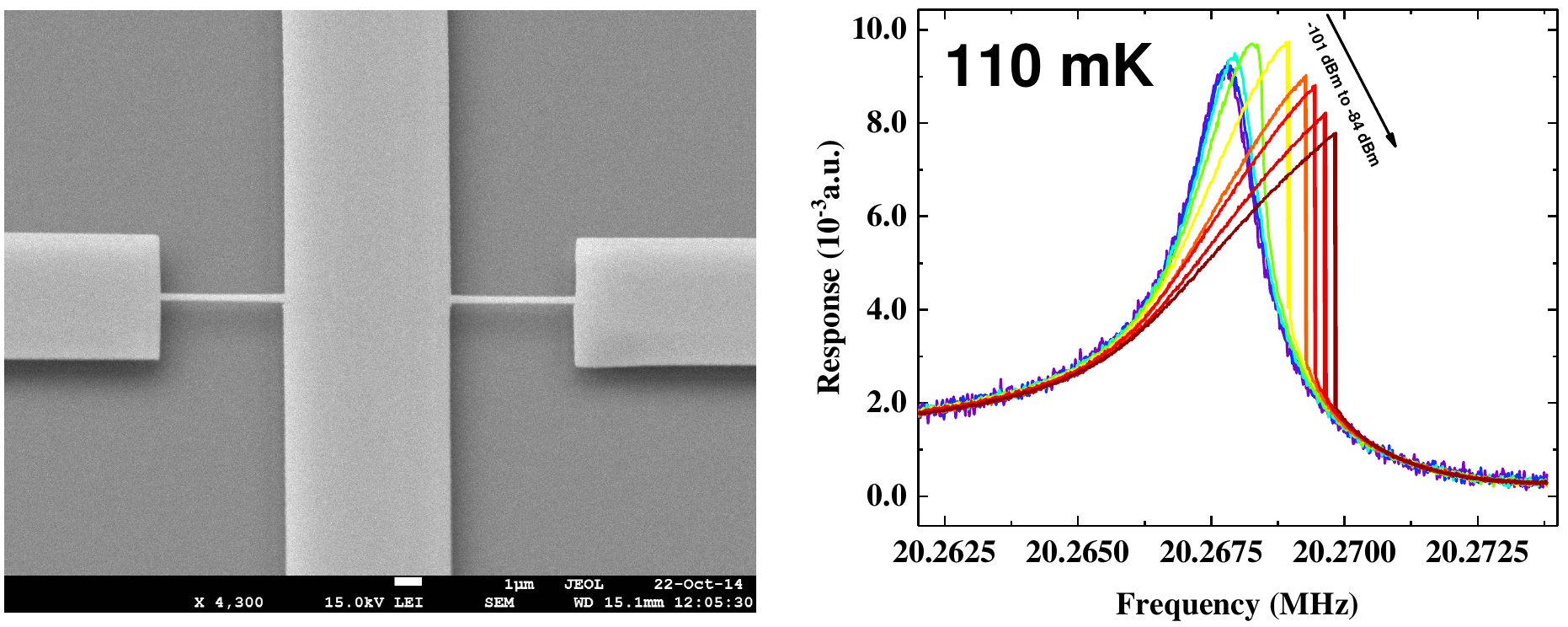}}
	\par
}\end{abstract}

\section{Introduction:}
Non-linear behaviour is ubiquitous in Nano-electro-mechanical systems (NEMS) due to the conducive aspect ratios (e.g. length to thickness) of nanoscale resonant structures and strongly coupled transduction schemes\cite{cross}. The fast response of these devices allows one to probe large domains of parameter spaces of dynamical phenomena like Arnold tongues\cite{arnold}. Some examples of non-linearity in NEMS include mechanical frequency mixers \cite{blick}, stochastic  amplifiers\cite{badzey}, non-linear inter-mode coupling phenomena \cite{lulla} and  nano-mechanical logic gates\cite{sinha}. Understanding dissipation in NEMS is critical for applications like sensors\cite{ekinci} signal processing\cite{dykman} and macroscopic quantum phenomena\cite{blencowe,poot}.  Non-linear oscillators are also better candidates to observe macroscopic quantum phenomena as it is easy to distinguish them from classical oscillator states due to uneven level spacing\cite{katz}. 

At temperatures typically below the boiling point of liquid helium where there is very little change of mechanical properties of materials the standard tunnelling model  phenomenologically maps  tunnelling two level systems (TLS) to various entities like defects, kinks and grain boundaries to explain dissipation in resonant mechanical systems as well as other resonant systems like electromagnetic cavities \cite{esqui}. TLS phenomena play a dominant role in low temperature dissipation in NEMS due to the enhanced surface to volume ratio\cite{mohanty_rev}.  Metallic NEMS devices of materials like Aluminium \cite{hakonen,hoehne} and Gold \cite{gold} have shown evidence of TLS dissipation scenarios at low temperatures. 

Palladium's affinity to hydrogen is well known. The adsorbed hydrogen occupies interstitial sites as $H+$ ions(protons) causing stress. In reference\cite{h2} stress due to adsorbed $H_2$ on nano-mechanical Au-Pd beams was used to sense hydrogen in ppm levels.  In a previous work \cite{av} on Pd nano-mechanical resonators in the  linear response regime  we showed tun-ability of TLS dissipation scenarios by exposure to very low pressures of $H_2$. The stress acting like a pseudo-Zeeman field on TLS allowed us to enhance the TLS phonon interactions. 

In contrast to the well established linear  damping proportional to velocity ($| {\bf f_{damp}} | = \gamma  v$ ) non-linear damping with cubic terms  formed by products of velocity and position( $|{\bf f_{nl}}| = \eta x^2 v $ or $|{\bf f_{nl}}| =\eta v^2 x $) has been recently observed in systems ranging from biological systems like cochlea of the ear to aeroplane structures and few NEMS devices \cite{eliott}. 
 Few nano/micro-mechanical systems like graphene beams, carbon nanotubes \cite{eichler} , diamond-resonators \cite{mohanty}, micromechanical Au-Pd beams \cite{buks} and graphene drums \cite{bockrath,vibhor} have demonstrated non-linear damping. Theoretical modelling of non-linear damping mechanisms are still at early stages with few systems like graphene having some plausible models based on intermodal coupling\cite{kinaret}. In this work we present response of Pd nano-mechanical resonators in the non-linear regime. We see evidence for non-linear damping that strongly depends on temperature from 110 mK to ~1.35  K where differential thermal expansion of devices is not expected to play any role( especially in top down fabricated devices). 
 
 \section{Non-linear damping Phenomena} 
 A Hookean harmonic oscillator with a potential $U(x) = \frac{kx^2}{2} $ and a frictional force proportional to the velocity i.e $|{ \bf f_{damp}}| =\gamma v$ 
can model a variety of systems ranging from molecules to large-scale engineering structures in a linear response regime\cite{pippard,baerlein}. For e.g., when a beam is driven hard and stretched far from its equilibrium length $l_{0}$ expanding the potential to higher order even terms $U(x) \sim \frac{kx^{2}}{2}  +  \frac{\alpha x^{4}}{4}$  gives rise to additional cubic restoring forces. Such systems are the classic Duffing oscillator.  The term $\alpha$ is usually small and produces a noticeable effect only at very large displacements\cite{baerlein,pippard,lakshmanan}. A simulation of a Duffing oscillator's  frequency response normalized to drive force (termed responsivity\cite{cross}) is shown in Fig.\ref{A}.A1 with characteristic features like frequency pulling ( shift to higher or lower values from linear regime), non-Lorentzian line shape and hysteresis depending on direction of frequency sweep. The responsivity peak remains same. Additional features like a phase portrait and the potential are shown in Fig\ref{A}.A2 \& A3.

Some theoretical works speculated the possibility of  an extra cubic non-linear damping term in a Duffing  oscillator\cite{mallik,cross} and even fractal powers\cite{fract}. The cubic damping term may be of the form $\eta x^{2} \dot{x}$ or $\eta x \dot{x}^{2}$. Irrespective of the form i.e, $x^{2}\dot{x}$ or $\dot{x}^2 x$ one can see as the amplitude or velocity increases the damping 
term increases in a non-linear fashion competing with the restoring forces. We use the form $\eta x^{2} \dot{x}$ in this work.  
 The equation of motion for a  non-linearly damped Duffing oscillator system is 
\begin{equation}
\label{nl}
m\ddot{x} + \gamma \dot{x} + k x +  \alpha x^{3}+\eta x^{2} \dot{x} = {F_{0} sin(\omega t)}  
\end{equation}

The frequency response of a cubic damped Duffing system simulated in Fig\ref{A}.A4. shows a non-Lorentzian line shape , but a drop in responsivity amplitude  and  absence of hysteresis for high ratios of $\eta/\alpha \sim 1.5 $. A transient phase plot is shown in Fig{\ref{A}.A5 exploring only one well. In analogy with centrifugal barriers used to model planetary motion \cite{baerlein} a pseudo-potential formed by the damping term is shown to depict the drop in turning points compared to the Duffing case.

\begin{figure}[ht]
	\centering
	\includegraphics[width=1\textwidth]{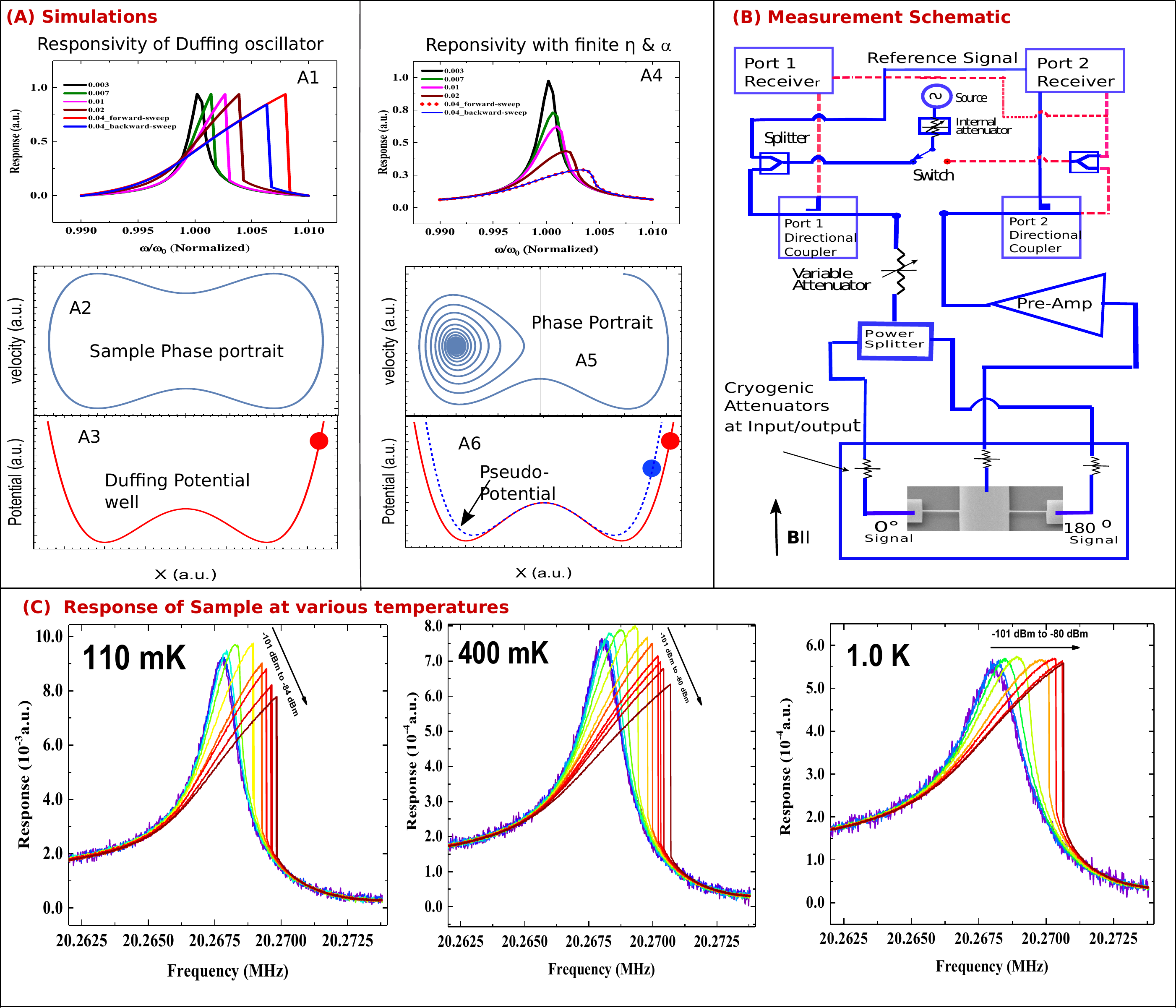}
		\caption{(A) {\bf Simulations of Duffing and non-linearly damped Duffing oscillators :} (A1)  Frequency response of a Duffing Oscillator with amplitude normalized to drive force showing frequency pulling and hysteresis with sweep direction. (A2) Phase portrait of a transiently forced Duffing oscillator showing it explores both wells. (A3) The double well potential of a Duffing system. The red ball represents a turning point for a given arbitrary Kinetic energy. (A4) Frequency response of  non-linearly damped Duffing oscillator for a ratio of $\eta/\alpha= 1.5 $ . Apart from frequency pulling it also shows 
			a dip in the normalized response amplitude with line broadening.Unlike Duffing hysteresis is absent(A5) Phase portrait for finite $\eta$ scenario showing it predominantly collapses into one well for a large transient force. (A6) Along with the Duffing potential in red the dotted blue curve represents a pseudo potential $\eta x^3\dot{x}$. This has a steeper barrier than the Duffing term. For the same initial kinetic energy the Duffing amplitude is represented by the red ball whereas the finite $\eta$ scenario has a lower turning point represented by the blue ball.
			(B) {\bf Measurement Schematic :}. Power from a Network Analyzer is sent via a variable attenuator and a 0 to 180 degree power splitter feeding two samples and output tapped from a common point to form a balanced RF bridge. Fixed attenuators at various input and output stages of the cryostat are used to minimize external heat loads and RF reflections and amplified by  50 ohm room temperature RF pre-amps. Once the variable attenuator is set to an optimal  range in the linear response regime of the device the power from port 1 of the Network Analyzer is varied. Port 2 measures S21 normalized to the output of port 1. The blue lines represent signal paths for $S_{21}$ parameters. The red lines are for $S_{12}$. The directional coupler along with internal switches allows to choose the measurement of $S_{21}$ with port 1 as a source and the internal power splitter allowing a comparison of response in the receiver at port 2.  
			(C){\bf Sample response at various temperatures:}.The linear regime collapses into one curve for all temperatures. The data shows a strong non-linear damping at a $T_{bath}\sim 110  \; mK $ indicated by a sharp drop in amplitude and a line broadening. There is an initial increase of amplitude only for temperatures below 1K when crossing the linear regime.  As the temperature approaches  $T_{bath}\sim 1.35\; K $ the amplitude drop and line broadening is less drastic compared to $T_{bath}\sim 110\; mK $. There is no hysteresis observed in any of this data sets with a 1Hz sweep step. }\label{A}
	
\end{figure}

\section{Measurement Scheme}
 We recapitulate the essential aspects of the measurement scheme in Fig\ref{A}.B. Detailed description of the nano-fabrication protocols of the samples and measurement schemes are given in reference\cite{av}. Radio frequency (RF) current from a vector network analyzer was driven through the sample with a magnetic field parallel to the wafer plane to excite and detect out-of-plane motion of the beam due to the Lorentz force. 
 The measured sample set B consisted of two samples sample B1 with dimensions a $\sim 4.35\mu\: m \times 390 \: nm \; (l \times w)$ and sample B2 with dimensions $\sim 4.5\mu m\: \times \: 366 \: nm \; (l \times w)$ both with 80nm thickness. The two samples in series  connected by $\sim 25$ micron wide input ports driven at opposite phases and a common out pad form a balanced RF bridge. The samples were soaked  to $\sim 2\times 10^{-3 }\: torr$ of $H_2$ while cooling down and subsequently pumped to lower $10^{-4} \:torr$
 when the mixing chamber temperature was below $\sim 160\: K$. The non-linear damping data for sample B1 which had a linear resonant frequency 
$f_0\sim 20.23 $ MHz was probed in detail at different temperatures. Another sample A1 from a set A without exposure to $H_2$  showed a weak non-linear damping at the lowest temperature.

At the lowest temperature the optimal power required to drive this sample to a non-linear regime was  $-93$dbm, whereas for samples with no exposure to $H_2$ was $ -80$ dbm  in an external magnetic field of 4 tesla. The quality factor (Q-factor) of the linear regime was $Q \sim 19500$ to $Q\sim 9800$ for the temperature range from $110\; mK$ to $1.35\;K$ . The overall power range applied to the resonators was from $-101\; dbm$ to $-80 \;dbm$ 
i.e., a maximum of $10 \; pW$. 
After fixing the value of some external attenuators (including a few at cryogenic stages to minimize reflections)  the drive power was varied  by the internal generator. The $S_{21}$ response is normalized to this drive power.  
In the linear regime the amplitude response curves extracted from $S_{21}$  collapse on top of each other except for some minor
difference due to noise at lower powers as shown in Fig(\ref{A})C ) at various drive powers for different temperatures. The induced voltage response was calibrated to the displacement in the linear regime(discussed in supplemental methods S1). An effective spring constant of $ k_{eff}\sim 129\; N/m$ for sample B1 and  $k_{eff}\sim 197.5\; N/m$ for sample A1 which was not exposed to hydrogen. 
The key features of the data in non-linear regime are discussed and analysed in the following section. 

\begin{figure}
	\centering
	
  	\includegraphics[width=0.9\textwidth]{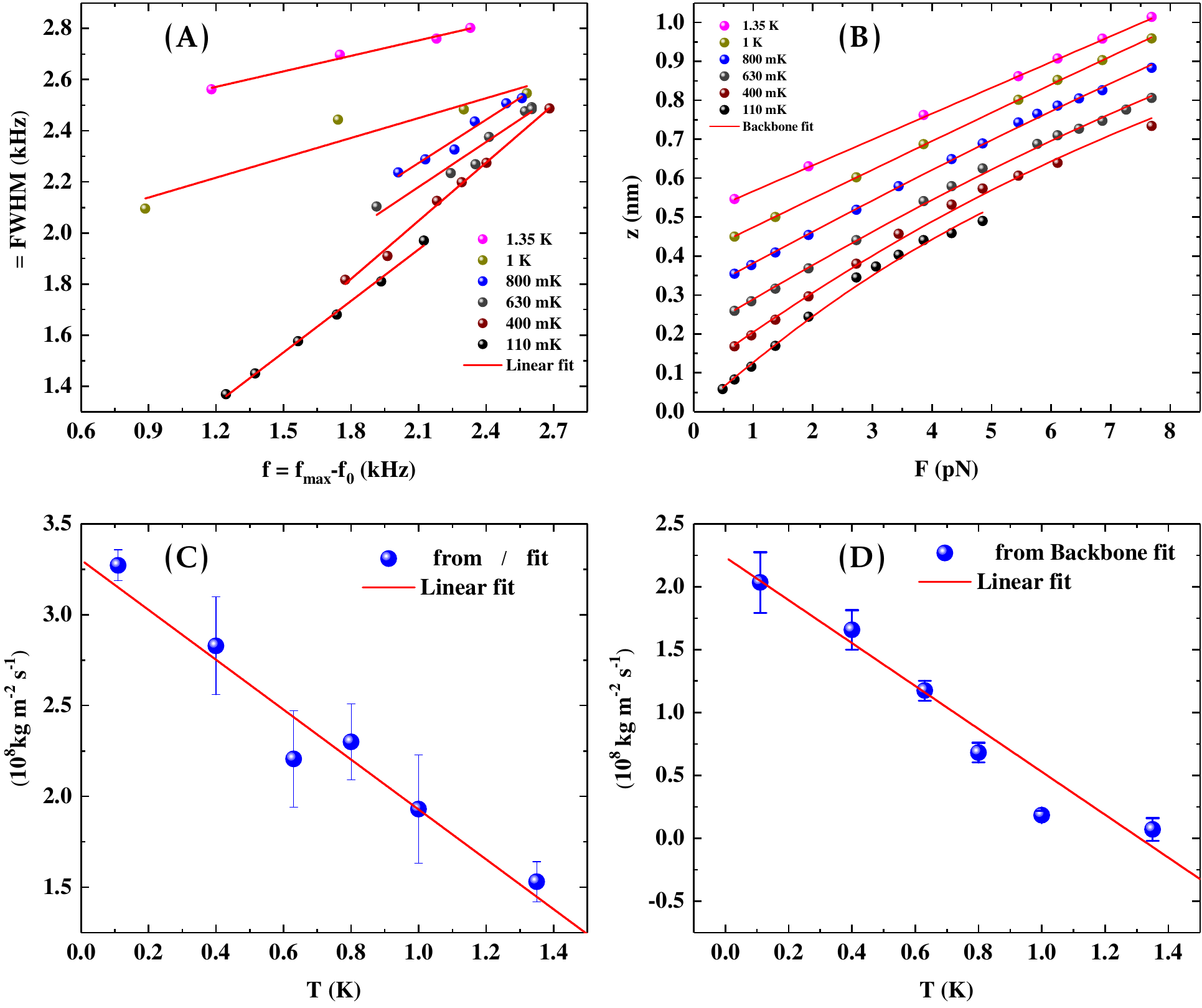}

		\caption{{\bf (a)} A plot of $\delta f$ vs $\Delta \Omega$ at various temperatures. The slopes are different showing a strong temperature dependence of $\eta/\alpha $. {\bf (b)} A plot of response peak rms amplitude vs drive force. The lower temperatures show a strong non-linearity. A fit to equation(2) with the frequency dependent term in the denominator taken to be zero. The algebraic expression is given in the supplementary materials. The curves are shifted by $0.1 \; pm$ for clarity with the linear terms also removed. 
		{\bf (c) } A linear fit to $\eta$ at different temperatures obtained by multiplying slopes of plot(a) with appropriate  Duffing constant$\alpha $ from back-bone fits to frequency pulling 
		according to equation (3).{\bf(d)}   A linear fit to $\eta$ obtained at different temperatures from plot (b). 
	}\label{res}
\end{figure}

\section{Results \& Discussions} 
The typical normalized response of resonator B1 is shown in Fig(\ref{A}.C). The onset of non-linearity shows a positive frequency pulling indicating that the resonator is effectively under some tensile stress. In the lower temperatures we see a slight increase in amplitude ( above the peak of the normalized linear $S_{21}$) when crossing the limits of linear response. This phenomena is less pronounced for $T \sim 630 \; mK $ and absent from $1 \; K$. An enhancement of Q-factor with drive power is observed in TLS dissipation scenario\cite{Phillips,cpwtls}. 
An ensemble of TLS responding to  the drive fields(electromagnetic or mechanical depending on the system) and saturating at higher powers is the reason. 
Despite the increase of amplitude there is a small broadening of the line-width keeping the effective Q-factor similar to the linear regime. In our devices there may be two competing mechanisms of dissipation, namely TLS saturation enhancing Q-factor, as well as a non-linearly damped Duffing oscillator regime. In contrast other  systems like surface acoustic waves \cite{sawtls} and quartz resonators \cite{quartz} are still in linear response regime and show an enhanced Q-factor.

At higher drive powers we see clear evidence for Duffing non-linearity and non-linear damping indicated by both a drop in amplitude and a broadening of the line-width. At higher temperatures the line broadening is more significant than the amplitude drop.

We analyse the data within the scope of secular perturbation theory of equation \ref{nl} following ref\cite{cross}. 
For a resonator of linear resonant frequency $\omega_0 =\sqrt{k/m_{eff}}$ and linear quality factor $Q$, the square of the amplitude response $z_0^2$ for a drive force with frequency $\omega$ and external force $F$ is given by
\begin{equation}
\label{secular}
z_0^2 = \frac{\left(\frac{F}{2m_{eff}\;\omega_0^2}\right)^2}{\left ( \frac{\omega-\omega_0}{\omega_0} -\frac{3}{8}\frac{\alpha}{m_{eff}\;\omega_0^2}z_0^2\right)^2 + \left(  \frac{1}{2} Q^{-1} + \frac{1}{8} \frac{\eta}{m_{eff}\;\omega_0}   z_0^2
\right)^2 } 
\end{equation}

 The term $\Delta \Omega = \omega - \omega_0 $ is the frequency pulling in the non-linear regime and $\delta f $ is the effective line width of the full width half maximum (FWHM) or 3dB points in the power spectrum of the device ( $ \delta f = Q^{-1 }f_0 $ for the linear regime).  From equation(\ref{secular})  it is obvious that the resonance occurs when the first term in the denominator goes to zero giving an expression for $\Delta \Omega $ as
\begin{equation}
\Delta \Omega = \frac{3}{8} \frac{\alpha}{m_{eff}\;\omega_0}z_0^2 
\end{equation}
This is also the standard back-bone curve for a Duffing oscillator without any cubic damping terms\cite{lakshmanan,cross} to obtain $\alpha$. 
Following \cite{cross} the effective damping term is $ \gamma_{eff} = \gamma + \frac{1}{4} \eta z_0^2 $ . In terms of $\gamma_{eff}$ the line width $\delta f$  of FWHM in the non-linear regime can be expressed as
\begin{equation}
\delta f = \frac{\gamma}{2\pi m_{eff}} + \frac{1}{8\pi m_{eff}} \eta z_0^2 
\end{equation}
The absence of hysteresis in our data prompted us use an approximation in the limit $ \gamma \rightarrow 0$\cite{eichler} to estimate $\eta$. Despite the reasonable fits, the critical ratio  $\frac{\eta}{\alpha} >\frac{\sqrt{3}}{2\pi}f_0 $ to kill bi-stability was not observed. The absence of hysteresis may be due to slight Euler buckling of our beams on exposure to $H_2$, although we cannot know the exact state of our beams at cryogenic temperatures. Samples cooled down with low pressures of  $H_2$ and also in vacuum showed slight buckling on imaging after a thermal cycling to room temperature. 

Equations (3) and (4) imply a plot of $\Delta \Omega \textrm{ vs } \delta f $ is linear\cite{bockrath,cross}  with a slope of $  \frac{2f_0}{3}\left(\frac{\eta}{\alpha}\right)  $ . The resonance peak and squared response was used to estimate the 3 dB points for this analysis. Since the frequency shift follows the classic back-bone curve for a standard Duffing oscillator, the peak in amplitude response happens when the first term in denominator of equation(\ref{secular}) is zero. Hence tracking the peak amplitude as a function of drive force gives an analog of the back bone curve for estimating $\eta$ \cite{mohanty}. The solution is given in supplementary section S3. 
We use both methods to estimate $\eta$ and the Duffing constant $\alpha$ from the standard back-bone curve. 

Both the line-width analysis Fig\ref{res}.A and the  amplitude method Fig\ref{res}.B show a similar inverse linear  trend in the temperature dependence of $\eta$ shown in  Fig(\ref{res}.C\& D. A fit up-to 1.35  K with an intercept or zero temperature damping of $3.3\times10^8 kg\;m^{-2}\;s^{-1}$ and slope of$   -1.37\times 10^8 \;kg\;m^{-2}\;s^{-1}/K $ is given as a guide in Fig\ref{res}.C. 
In Fig \ref{res}.D the fit with an intercept of $2.2\times10^8 kg\;m^{-2}\;s^{-1}$ and slope of  $ -1.70 \times 10^8 \;kg\;m^{-2}\;s^{-1}/K $ is given as a guide. As the amplitude reduces at higher temperatures the estimates saturate in this technique whereas the line width was still a better indicator of the presence of non-linear damping. The error bars for the last two points in Fig.2D  are too small to see in this scale. We can state that within the limits of  estimated error bars from fits both methods give a similar order of magnitude of $\eta$ and a trend of linear drop with increasing temperature. 

 The Duffing constant $\alpha $ was non-monotonic with temperature in the range of $\alpha \sim 0.92 - 1.9 \times 10^{17}kg\;m^{-2}\;s^{-2}$ . A back-bone fit for estimating $\alpha$ is given in the supplementary material S2. The Duffing constant $\alpha$ for sample A1, cooled in vacuum, is an order of magnitude less with $\alpha \sim 9.5 \times 10^{15} kg\;m^{-2}\;s^{-2}$ . Thus we have managed to enhance the non-linear elastic constant with exposure to $H_2$. The estimated non-linear damping $\eta$ for A1 cooled in vacuum  was $\eta \sim 2.5 \times 10^7 \;kg\;m^{-2}\;s^{-1}$ at $T\sim 160\; mK$, showing it is lower by an order of magnitude. 

Our data shows a clear drop in the damping parameter $\eta$ with increasing temperature. No other NEMS systems have shown any temperature dependence of non-linear damping in these temperature ranges. 
Systems like carbon  nanotubes or graphene beams in ref\cite{eichler} showed dependence of non-linear damping $\eta$ on the tensile state of the beam. A carbon nano tube was reported to have similar $\eta$ at  $400 \; mK$ and $ 5 K$   speculating that this may be due to Van-der-Waals forces at the clamping points. In diamond NEMS non-linear damping was seen below $77 \;K$ with no significant change in $\eta$ when cooled down to 55mK\cite{mohanty}. 

We can exclude simple joule heating as the applied power is very low and our system is a conductive monolithic metallic beam connected to large micron scale pads. Simple joule heating is also expected to cause frequency shifts due to diffusion of adsorbed $H_2$,  as even ppm levels of $H_2$  adsorption was sensed in ref\cite{h2} with lower Q-factors at room temperature, hence any mass redistribution will be signalled by a jump in frequency as well as other diffusion induced bi-stability phenomena\cite{atalaya}.  We find no hysteresis in sweeps of forward and reverse directions implying joule heating is negligible at these powers.  

A recent theory by Atalaya et.al \cite{dykmanNL} proposed several mechanisms involving flexural modes exciting thermal two-phonon scattering processes within a NEMS or MEMS device causing non-linear damping in various scenarios. 
The non-linear thermoelastic-damping they predict is expected to increase $\eta$ with temperature. 
 
 Reference\cite{dykmanNL} also proposed a non-linear analogue of the Akhiezer mechanism. The Akhiezer mechanism involves phonons coupling to strain fields that oscillate faster than thermal phonon relaxation rates for the system. The individual phonon modes attain a different temperature and relax to the bath temperature (equilibrium temperature of the whole beam in this case). The heat flow between different phonon modes leads to entropy production and damping\cite{aluru}. Unlike $I^2 R$ heat flow which is a $2f$ harmonic of the drive term, this process for the non-linear case is a sub harmonic, as strain fields vary faster and need not result in drastic thermal gradients but mere fluctuations around a mean.
   
The criterion to observe non-linear Akhiezer damping requires the device angular frequency $\omega_0 >> \frac{v_sl_{phonon}}{L^2} $ (where speed of sound $v_s =\sqrt{(E/\rho)}\sim 3100\; m/s$ with Young's modulus $E\sim 120\; GPa $  of Pd and $\rho$ density ) , $
l_{phonon}$ is the mean free path of phonons and $L$ is the device length. If we assume $l_{phonon} \sim t$ device thickness the estimated minimum  frequency is $\frac{\omega_0}{2\pi} >> 2 \; MHz $ and we are at $ \frac{\omega_0}{2\pi}\sim  20 \: MHz$ the criterion is clearly satisfied.  There is no thermal conductivity data on Pd films. Data for AuPd films reported $l_{phonon}\sim 25\;nm $ for temperatures around $1K$ \cite{wybourne}. It is plausible to expect that pure $Pd$ being a pure metal may have slightly longer mean free paths comparable to the thickness and assuming the thickness of the beam as the limiting factor is reasonable.
 Also the effectively  lower spring constant for beams exposed to $H_2$ implies an effectively lower Young's Modulus($E$) as opposed to beams cooled in vacuum. Hence  the speed of sound ($v_s\sim \sqrt{E/\rho} $) in these estimates can also be reduced, further enhancing the criterion to observe Akhiezer damping. 

An anisotropic Gr\"{u}neisen parameter is expected to play a role in the non-linear Akhiezer damping . We do not have a direct measure of the Gr\"{u}neisen coupling for the acoustic phonon modes. The literature reports an enhanced Gr\"{u}neisen constant from 2 to 3 for bulk Pd  and $\alpha$-hydride state i.e., less than $Pd_{1-x}H_x\;x<0.6$ \cite{grun}. We have not electrolytically loaded our samples with $H_2$ but with the aspect ratio of our sub-micron devices even a small dose can have a drastic effect.  Despite the phenomenological inference of enhanced phonon-TLS coupling in our linear data \cite{av} it also points to  plausible enhancement of Gr\"{u}neissen parameters.

 Overall there are several mechanisms that contribute to dissipation. In linear regime the TLS mechanism is dominant. While other mechanisms induced by clamping and  eddy currents may add to the TLS background limiting the Q-factors. In the non-linear regime the TLS will get excited beyond two levels and need not be restricted to independent two level systems\cite{strain}. We see phenomena similar to saturation of TLS at slightly higher powers with simultaneous onset of non-linear damping taking over as the dominant mechanism. The novel non-linear Akhiezer mechanism in reference\cite{dykmanNL} is the most plausible scenario. The predicted linear drop of $\eta$ has been verified by two independent analysis within the scope of secular perturbation theory. The enhancement of an anisotropic Gr\"{u}niesen parameter is a key requisite for this mechanism to survive. Palladium thin films are known to form larger grains of few tens of nm and may aid this anisotropy. The large grain size at higher deposition rates $\sim 0.2nm /s $ has been  used to fabricate reliable cryogenic resistors \cite{pdresistor}. Materials like gold did not report \cite{li} non-linear damping, possibly due to smaller grain sizes and also the absence of compressive strains we introduce by adding $H_2$. While we do not have an estimate of the Gr\"{u}neisen parameters, the softening of our beams on exposure to $H_2$ via compressive strain in Pd and a value of $\eta$ which is one order of magnitude less in vacuum cooled device is sufficient to say we have managed to affect the strain-field phonon coupling. We may expect a reduction of non-linear damping $\eta$ in devices of smaller grain size like AuPd and conversely enhancement in more oriented or epitaxial films. Enhancement of the Duffing constant $\alpha$ with reduction of non-linear damping $\eta$ in alloys may pave way for non-linear mechanical devices in the quantum regime\cite{katz}.

\begin{acknowledgement}
	We greatly benefited in getting a sound start on this work with access to the nano-fabrication facilities in the research group of  Prof D. Weiss at the University of Regensburg.  We thank Dr J. Eroms and Mrs C. Linz for their assistance. We also thank Prof D. Weiss for a thorough review of this manuscript. We thank Prof Sudeshna Sinha for guidance on simulations and several detailed discussions. We thank Prof A.D. Armour for reviewing this manuscript. We thank DST (India) Nanomission Project No.SR/NM/NS-1098/2011, DST (India) Ramanujan Fellowship
	Project No. SR/S2/RJN-26/2010, and IISER (India) at Mohali for funds. We also thank DST (India) for Inspire Fellowships
	and CSIR-UGC (India) for funding students. We thank the  SEM facility at IISER Mohali ( especially former member  Mr Inderjit Singh and current staff Mr.Vivek Singh for technical assistance.)

\end{acknowledgement}

\providecommand{\latin}[1]{#1}
\makeatletter
\providecommand{\doi}
{\begingroup\let\do\@makeother\dospecials
	\catcode`\{=1 \catcode`\}=2 \doi@aux}
\providecommand{\doi@aux}[1]{\endgroup\texttt{#1}}
\makeatother
\providecommand*\mcitethebibliography{\thebibliography}
\csname @ifundefined\endcsname{endmcitethebibliography}
{\let\endmcitethebibliography\endthebibliography}{}


\end{document}


\maketitle

\section*{S1. Calibration of beam displacement in Linear response regime}

The network analyzer measures ratio of power with respect to the drive power. The $S_{21}$ linear  amplitude response is converted to arbitrary power by squaring the signal. This is used to determine the line widths in the non-linear regime.  Factoring the squared $S_{21}$ with the reference power  we get the power. 
We convert this signal to $V_{rms}$ using $V_{rms} = \sqrt{{P}{Z}}$ with  $Z \sim 50\;\Omega $ as the low impedance device is terminated with $50\;\Omega $ attenuators . In linear response regime the rms displacement\cite{displ} at resonance is given by  $z_{max} = \frac{FQ}{K_{eff}} $ and the induced peak emf $ V_{peak } =  \frac{\xi |{\bf B^2}| l^2 I Q}{m\omega_0} $ where $F = ILB$ the drive force($I$ is estimated against a $50\; \Omega$ load for total drive power) , $K_{eff}$ effective spring constant, $l$ length of beam, $\xi = 0.81 $ is a mode shape constant for a clamped beam ,  ${\bf B}$ is the applied magnetic field. The effective displacement is given by 

\begin{equation}\label{displacement}
	z_{max} = \frac{V_{peak}}{\xi\;l |{\bf B}| \omega_{0}}\tag{1.S}
\end{equation}

\begin{figure}[ht]
	\renewcommand{\figurename}{Fig.S.}
	\centering
	\begin{minipage}[b]{0.45\textwidth}
		
		\begin{overpic}[width=1\textwidth]{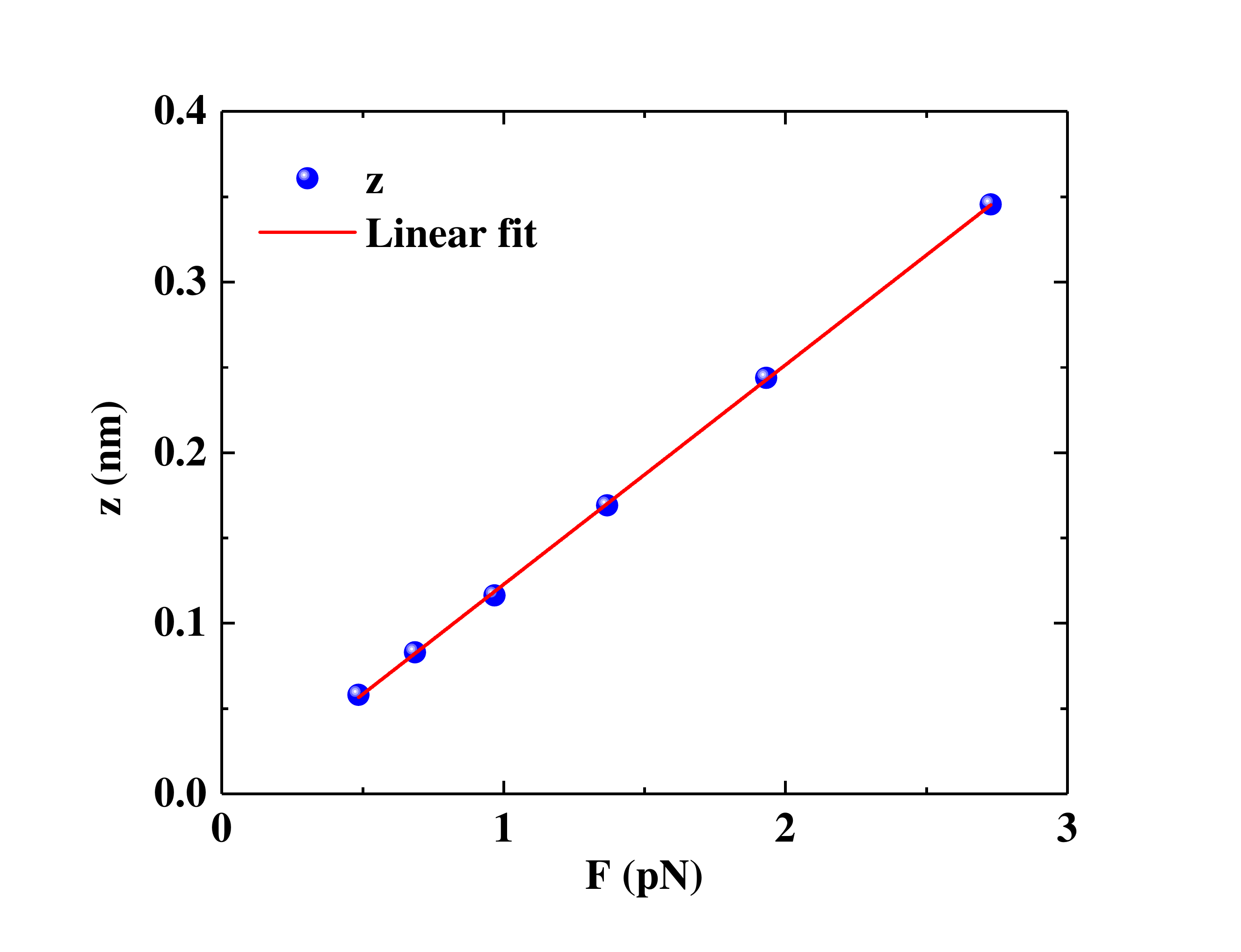}
			\put (20,40) {\large (A)}

	\end{overpic}

	\end{minipage}
	\hfill
	\begin{minipage}[b]{0.45\textwidth}
		 	\begin{overpic}[width=1\textwidth]{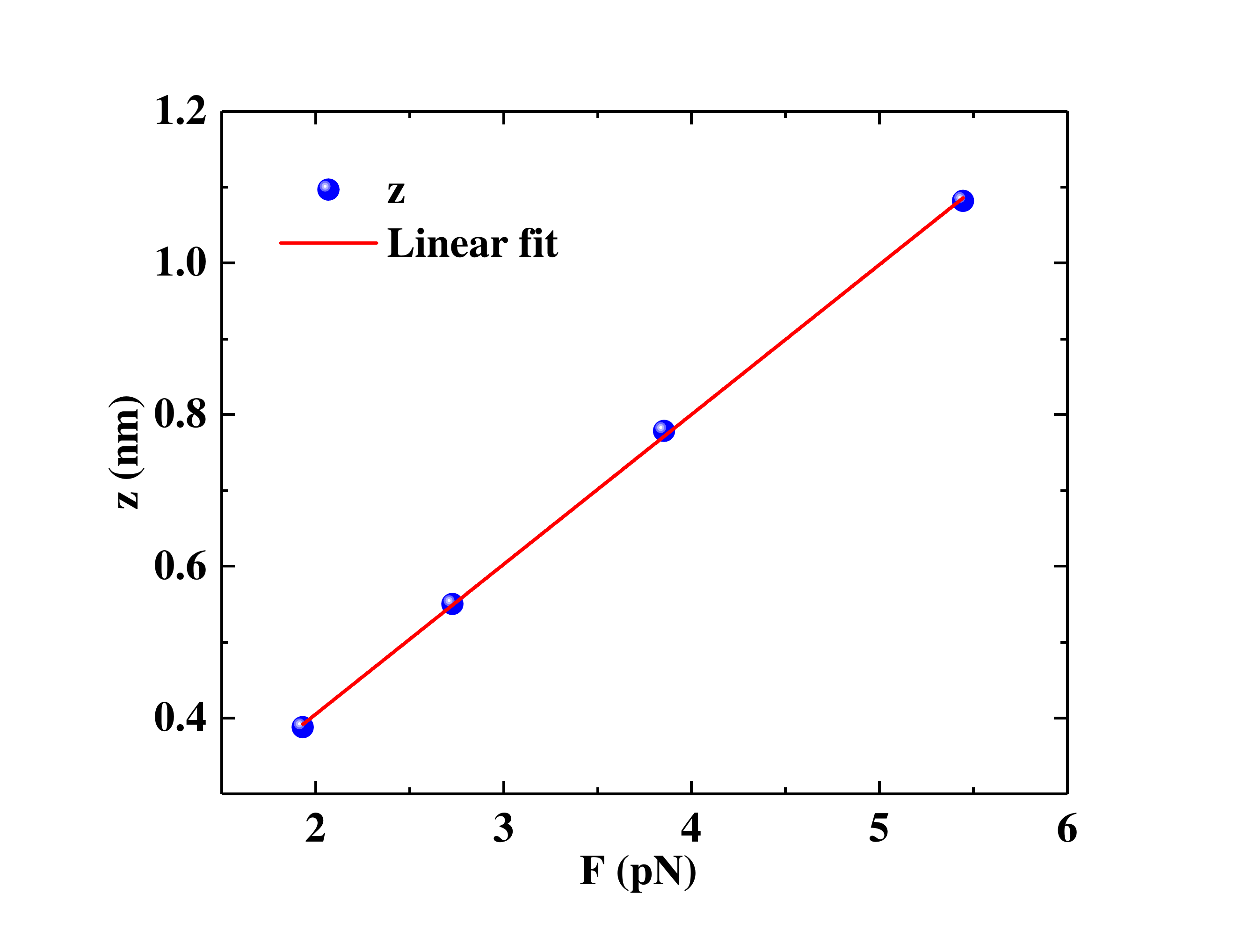}
		 	\put (20,40) {\large (B)}
		 	\end{overpic}

	\end{minipage}
\caption{\scriptsize (A) A plot of Force vs Displacement for hydrogen exposed sample B1 at 110 mK yielding an effective spring constant $K_{eff} 129 N/m $  (B) A plot Force vs Displacement for sample A1 cooled in vacuum at 160 mK yielding an effective spring constant $K_{eff}= 197.5 N/m $ }
\end{figure}

\begin{figure}
	\renewcommand{\figurename}{Fig.S}
\scalebox{0.5}{	\includegraphics{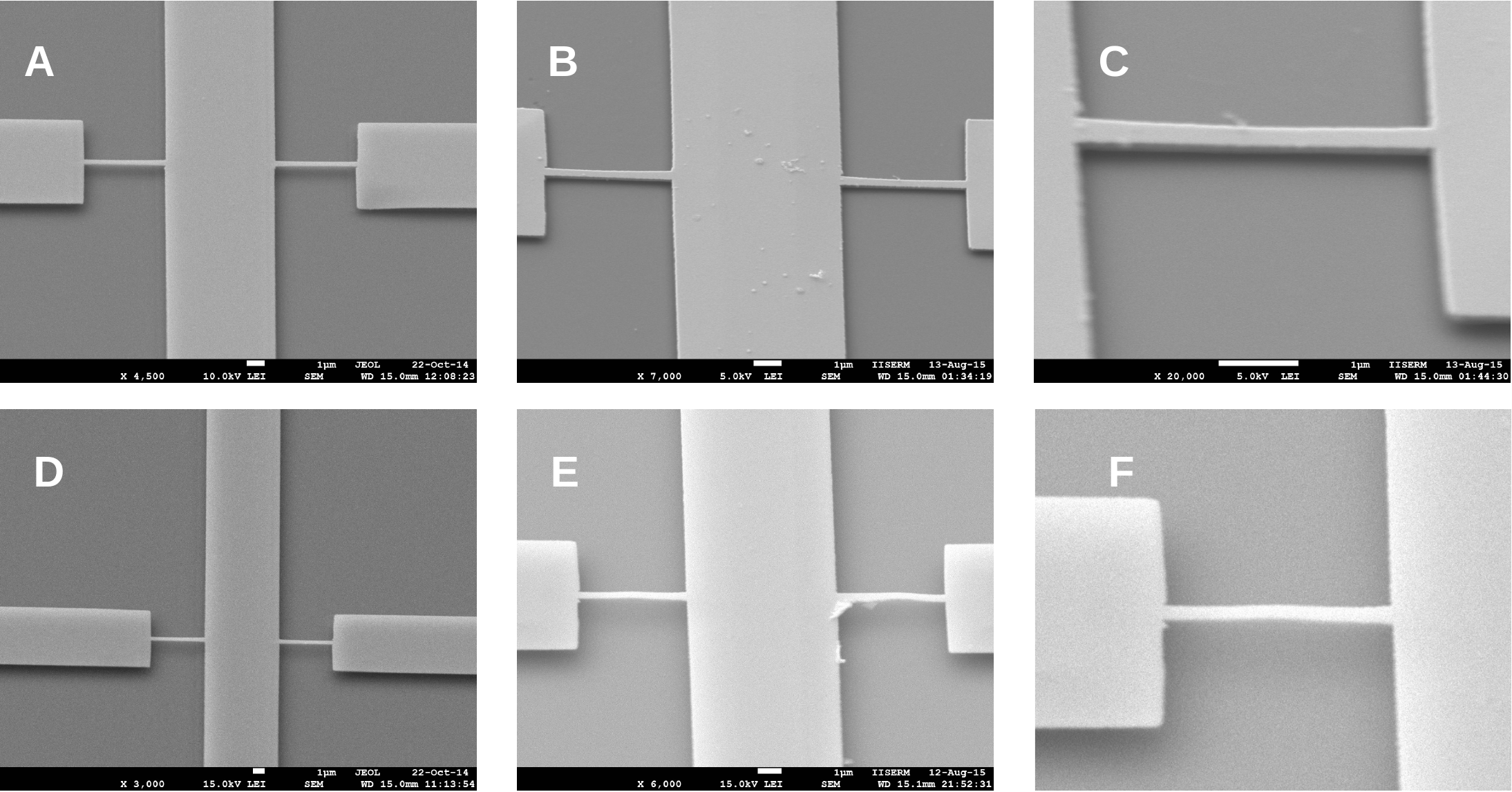}}
\caption{\scriptsize SEM micro-graphs of samples (A)Sample set B before measuring (B) Sample set B after measuring with two cool-downs in $H_2$ atmosphere. First with $\sim 1\times 10^-3\; torr $ and second in $\sim 1\times 10^-2\; torr $. There is a slight sagging of the beam after the cool-down.  (C)A zoomed view of one of the samples in B. (D) Sample set A which was measured only in vacuum. (E) Sample set A after cool-down in vacuum. Sample A2 had piece of Indium stuck under it. 
(F)Enlarged view of one beam showing a sagging. }\label{samples}
\end{figure}

The amplifier gave a factor of $67\;dB$ power gain under a bias condition of 12 Volts with short hand form-able low loss copper coaxial cables connected. The actual gain  factor was lower due to small attenuators and cupro-nickel cables in the cryostat.  A plot of displacement vs force was found to have an intercept one order of magnitude smaller than the smallest displacement. A corrected gain by 1 dB from the initial estimate was around $62.5\;dB$ was used the calibrate the voltage and displacement in linear regime to obtain a zero intercept. These estimates can be off by small values  2 to 3 dB as it is hard to know the exact attenuation at cryogenic conditions using very low powers. In the non-linear regime the calibration of the induced voltage was used as to obtain the effective displacement. 
 Also the mode constant $\xi$ is not exactly known. Fig(2) shows images of samples before and after measurements. Both samples exposed to $H_2$ and samples cooled in vacuum show some minor buckling after bringing the system to room temperature. 
We have no means of knowing the true status of the beam at cryogenic  temperatures. 
Despite the small errors in these estimates the overall estimates are self consistent within some errors. 

\section*{S2 Estimation of $\alpha$ from back-bone curve}

\begin{figure}[H]
	\centering
		\renewcommand{\figurename}{Fig.S}
	\includegraphics[width=0.95\linewidth]{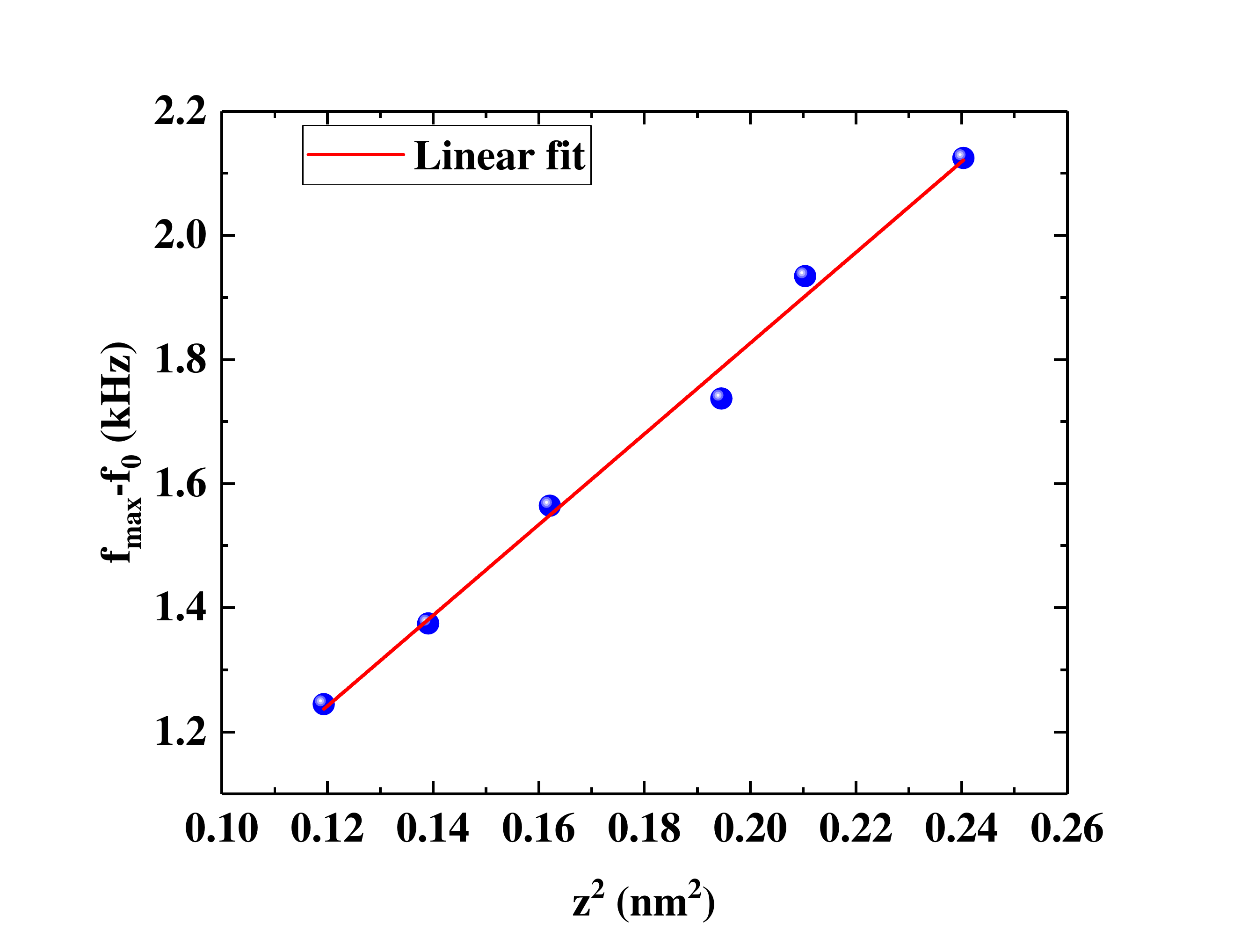}
	\caption{A plot of frequency pulling from linear response frequency($f_0$) against the squared amplitude $z_0^2$ to estimate the Duffing constant $\alpha$ for sample B1 at $T\sim 110mK$ .We get an $\alpha\sim 1.24 \times 10^{17}N/m^2$  }
	\label{a}

\end{figure}

As described in the main text the response in secular perturbation approach has a peak when the first term in the denominator of equation(2)is zero. The solution follows the standard back-bone curve for the Duffing oscillator is given by 

\begin{equation}\label{alpha}
\omega_{max} =\omega_{0}+\frac{3}{8}\frac{\alpha}{m_{eff}\;\omega_0}({ z_0 }^{2}_{max})\tag{2.S}
\end{equation}
The term $\alpha$ is like a cubic elastic constant. A positive $\alpha$ indicates a stiffening response to the drive force seen by shift to higher frequencies as seen in our data. A negative $\alpha$ implies a softening response to the drive force. A plot of squared amplitude linearizes the above equation. A sample data showing frequency shift peaks a plot of frequency pulling vs  squared amplitude is shown in Fig(\ref{a}) for sample B1 at $T\sim 110 mK$ .

As mentioned in the main text $\alpha$ was non-monotonic with temperature with values ranging from $0.92 -1.91 \times 10^{17}N/m^2$ . This  indicates that once some elastic properties are fixed it is of  geometric nature. 
The Duffing constant for a beam cooled in vacuum is $\alpha \sim 9.5 \times 10^{15}N/m^2$ indicating it is atleast an order of magnitude smaller. As mentioned in the text alloys may allow us to engineer higher $\alpha$ with lower damping $\eta$ 

\section*{S3. Estimation of $\eta$ from displacement}
The response of the beam within the scope of secular perturbation theory is given by equation(2) in the text. 
\begin{equation}
\label{secular}
z_0^2 = \frac{\left(\frac{F}{2m_{eff}\;\omega_0^2}\right)^2}{\left ( \frac{\omega-\omega_0}{\omega_0} -\frac{3}{8}\frac{\alpha}{m_{eff}\;\omega_0^2}z_0^2\right)^2 + \left(  \frac{1}{2} Q^{-1} + \frac{1}{8} \frac{\eta}{m_{eff}\;\omega_0}   z_0^2
	\right)^2 } \tag{3.S}
\end{equation}
As mentioned in the text at resonance the frequency term in the denominator is zero. Solving
the response  algebraically using a  symbolic package like Mathematica we get an expressions for 
$z_{max}$. The imaginary roots are and the solution that gives a  negative amplitude for a given $\eta $ , drive force and $Q$ are discarded. The physically plausible solution  is  given below

\begin{equation}\label{eta}
\begin{aligned}
z_{max}= {} & \ \frac{\sqrt[3]{\frac{32}{3}}k_0Q^{-1}}{\left((-9F\eta^{2}\omega_0^{2}+\sqrt{3}\sqrt({16Q^{-3}k_0^{3}}\eta^{3}\omega_0^{3}+27F^{2}\eta^{4}\omega_0^{4})\right)^{\frac{1}{3}}} \\  
& \ -\sqrt[3]{\frac{2}{9}}\frac{{\left((-9F\eta^{2}\omega_0^{2}+\sqrt{3}\sqrt({16Q^{-3}k_0^{3}}\eta^{3}\omega_0^{3}+27F^{2}\eta^{4}\omega_0^{4})\right)^{\frac{1}{3}}}}{\eta\omega_0}  
\end{aligned}\tag{4.S}
\end{equation}

The normalized signal can be converted to a voltage and in-turn to a displacement amplitude as described in previous section. Plotting the amplitude vs the drive force and using trial values for $\eta$ we can estimate from a fit to equation(\ref{eta}).

 \section*{S3 Analysis of Non-Hydrogenated sample at lowest temperature}
 
We were able to collect the data for a sample cooled in vacuum that displayed a higher linear spring constant and a weaker non-linear damping.  
A fit of line width vs frequency shift\cite{Lifshitz,Bockrath} as discussed in the main text is given in Fig{\ref{lw} 
 
 \begin{figure}[H]
 \renewcommand{\figurename}{Fig.S}
 	\centering
 	\includegraphics[width=0.95\linewidth]{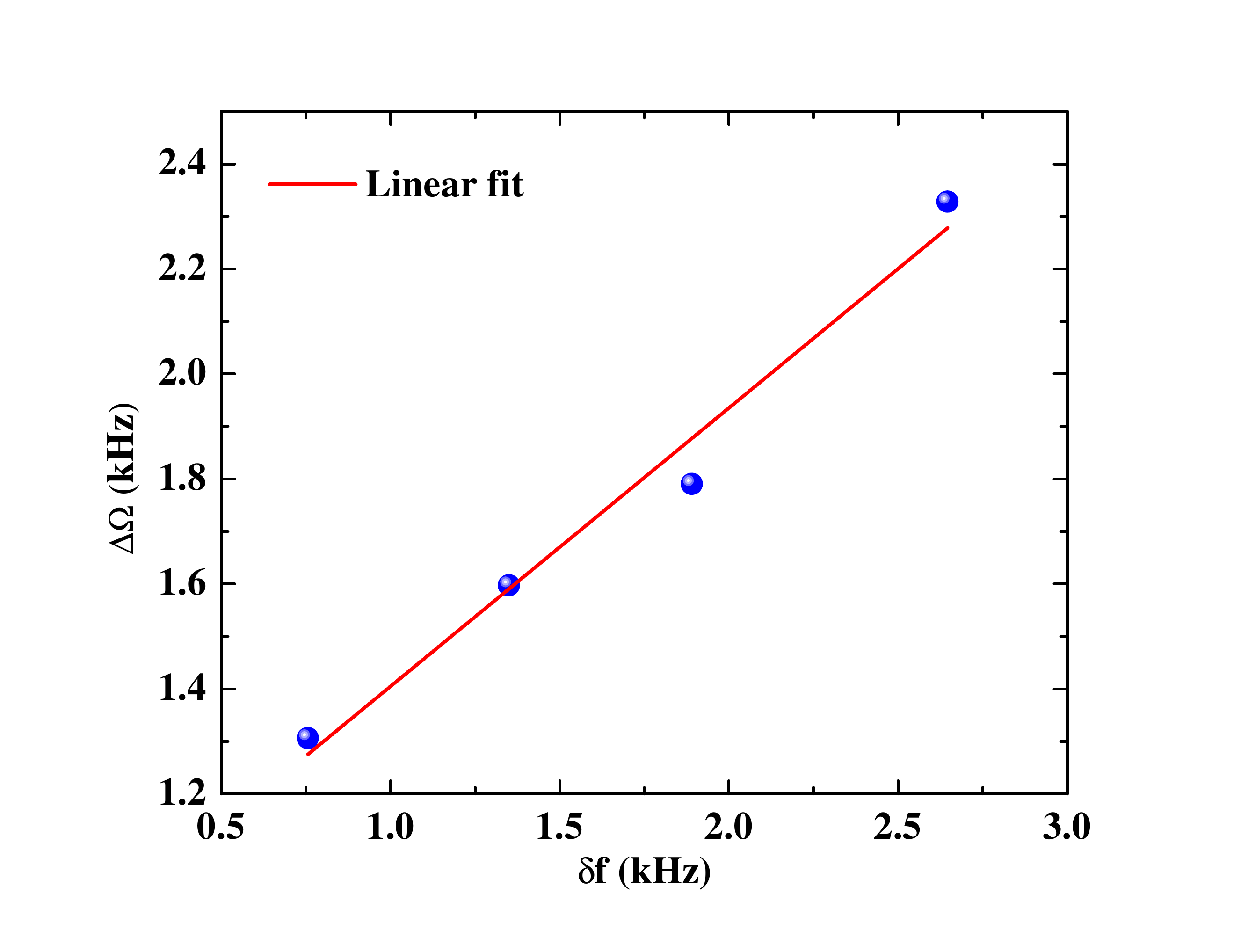}
 	\caption{Line width vs frequency pulling plot for  }
 	\label{lw}
 \end{figure}

Value of $\eta$ by dividing this slope with $\alpha$ yields    $\eta \sim 2.148*10^{7}$ $kgm^{-2}s^{-1}$. This is one order of magnitude smaller than the sample exposed to$H-2$ at such low temperatures. 

The amplitude analysis as discussed in main text as the back-bone analog \cite{mohanty} using equation(\ref{eta})  to estimate $\eta$  for the same sample is given below if Fig.S.\ref{amppd2c3}. This method does not inherently depend on $\alpha$. It must be noted that the effective mass of the resonator from $\omega =\sqrt{k/m_{eff}}$ gives the correct value of $\alpha$ rather than using the estimated mass from volume of the beam.vElse the estimates from the line width and amplitude discussed below are different. The enhanced effective mass could be due to some parts of the undercut supporting fins are also driven.

\begin{figure}[H]
	\renewcommand{\figurename}{Fig.S}
	\centering
	\includegraphics[width=0.95\linewidth]{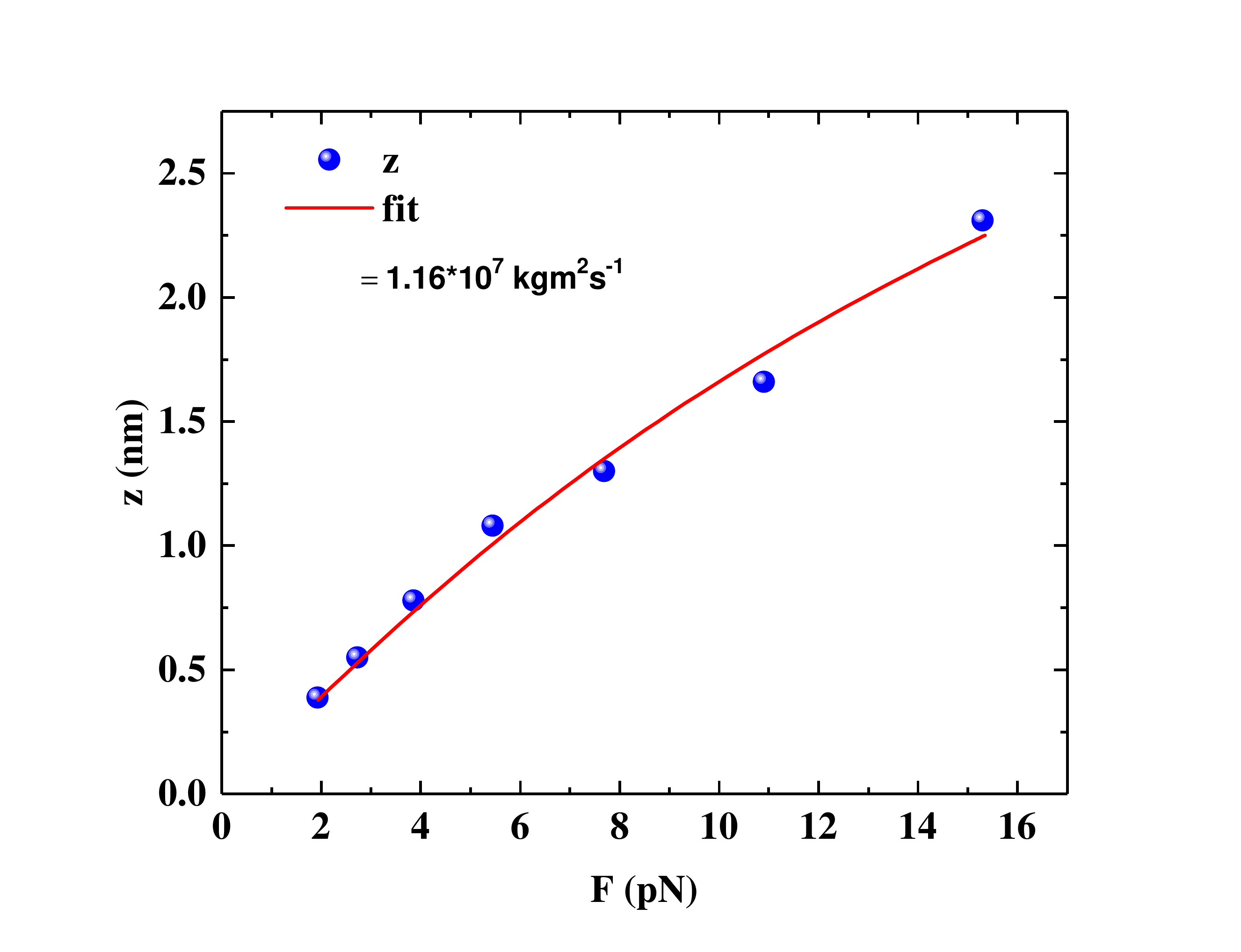}
	\caption{Fitting of non hydrogenated sample by Backbone curve }
	\label{amppd2c3}
\end{figure}

Value of $\eta$ by $\eta$/ $\alpha$ analysis is $1.16*10^{7}$ $kgm^{-2}s^{-1}$. Overall both estimates give values that are an order of magnitude less at these temperatures.

\section*{S4 Absence of hysteresis }

Even with the lowest step size of 1 Hz  we did not see any hysteresis in the non-linear response with a sweep approaching the resonance in a forward and reverse direction. A sample data set is shown in Figure(\ref{hyst}) for sample B1 at 110 mK 

\begin{figure}[H]
	\renewcommand{\figurename}{Fig.S}
	\centering
	\includegraphics[width=0.95\linewidth]{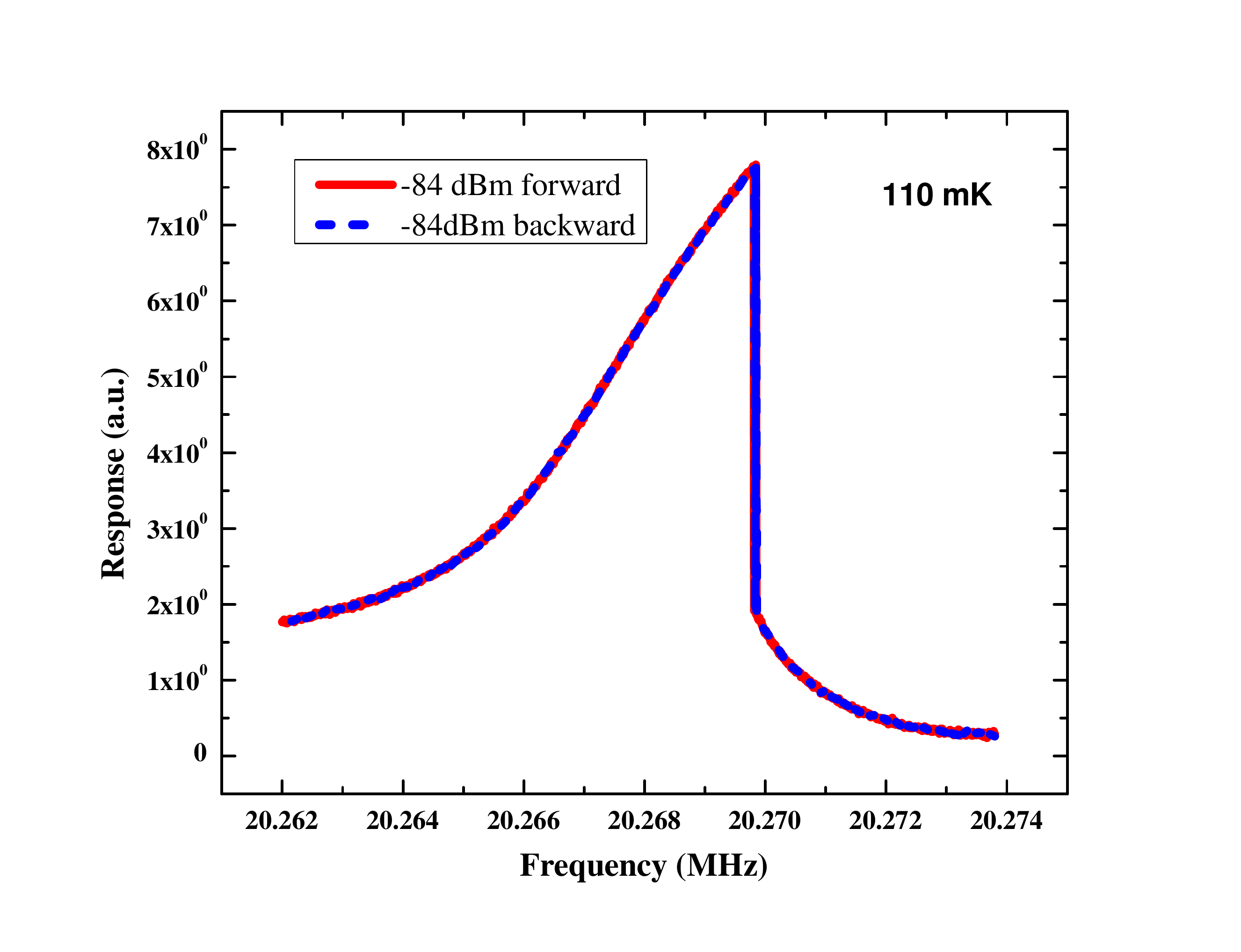}
	\caption{The red curve is forward and the blue reverse swee, Both curves overlap. }
	\label{hyst}
\end{figure}

While absence of hysteresis might indicate a critical magnitude of $eta$ to kill bi stability.

\section*{S5 Grain Size}

We found that under evaporation conditions of $1.5-2.0 \;$ \AA/ s  thin film deposition rate for the devices $Pd$ grains were much more prominent than $Au$ for similar rates. This could be due to higher melting point of $Pd$ gives it more kinetic energy as well as nucleation conditions of the substrate. While typical sizes of grains may be $\sim 10\; nm $ are visible in both films, the amount of oriented grains seems very high in $Pd$. Gold with a higher atomic number is expected to give a better contrast due to a Z-factor induced enhancement of scattering cross section for electrons. We found in  $Pd$ films and devices it was easier to focus the grains as it clearly had more oriented grains than gold. Figure\ref{grain} shows SEM micrographs of Palladium and Gold films. 

\begin{figure}
	\renewcommand{\figurename}{Fig.S}
	\begin{minipage}[b]{0.45\textwidth}
		\begin{overpic}[width=1\textwidth]{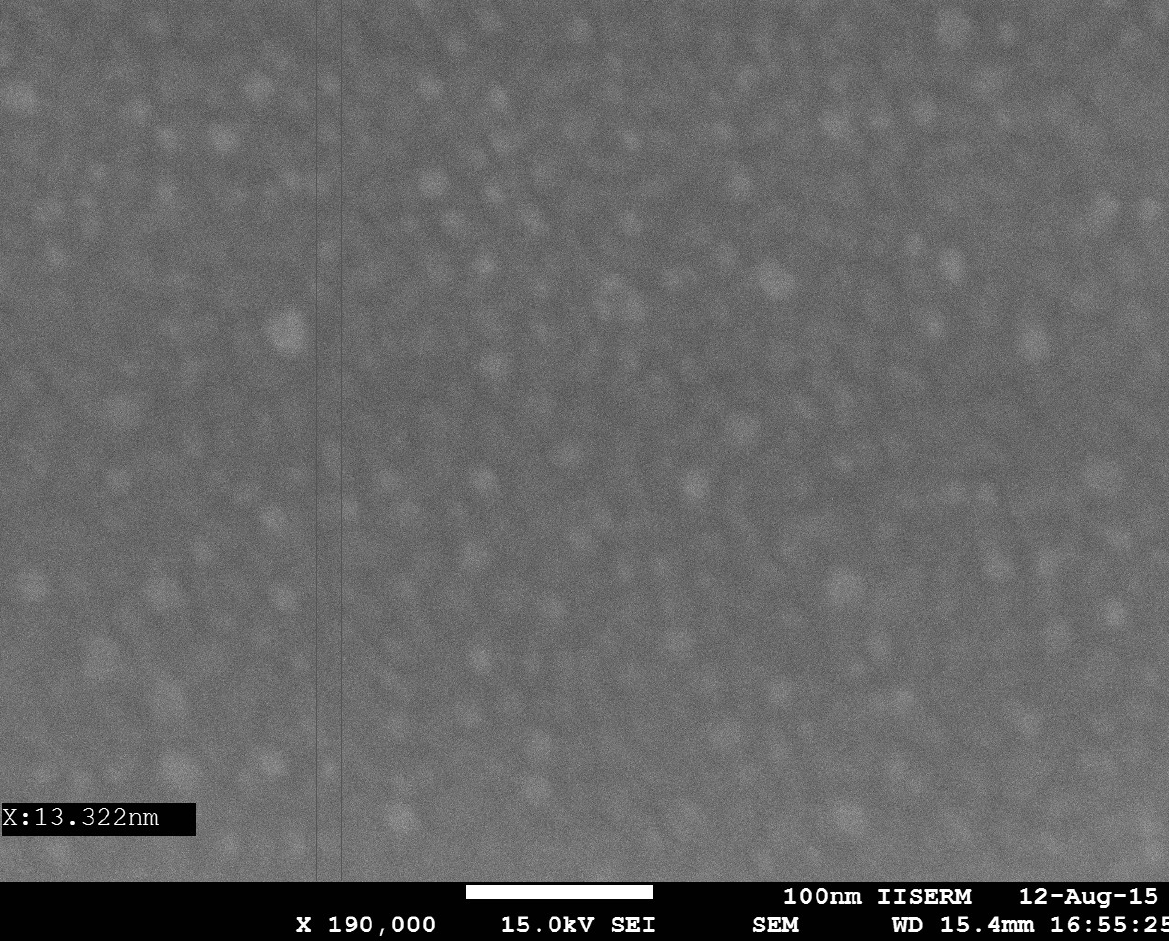}
		\put (5,68) {\large (A)}
		\end{overpic}
\end{minipage}
\hfill
	\begin{minipage}[b]{0.45\textwidth}
		\begin{overpic}[width=1\textwidth]{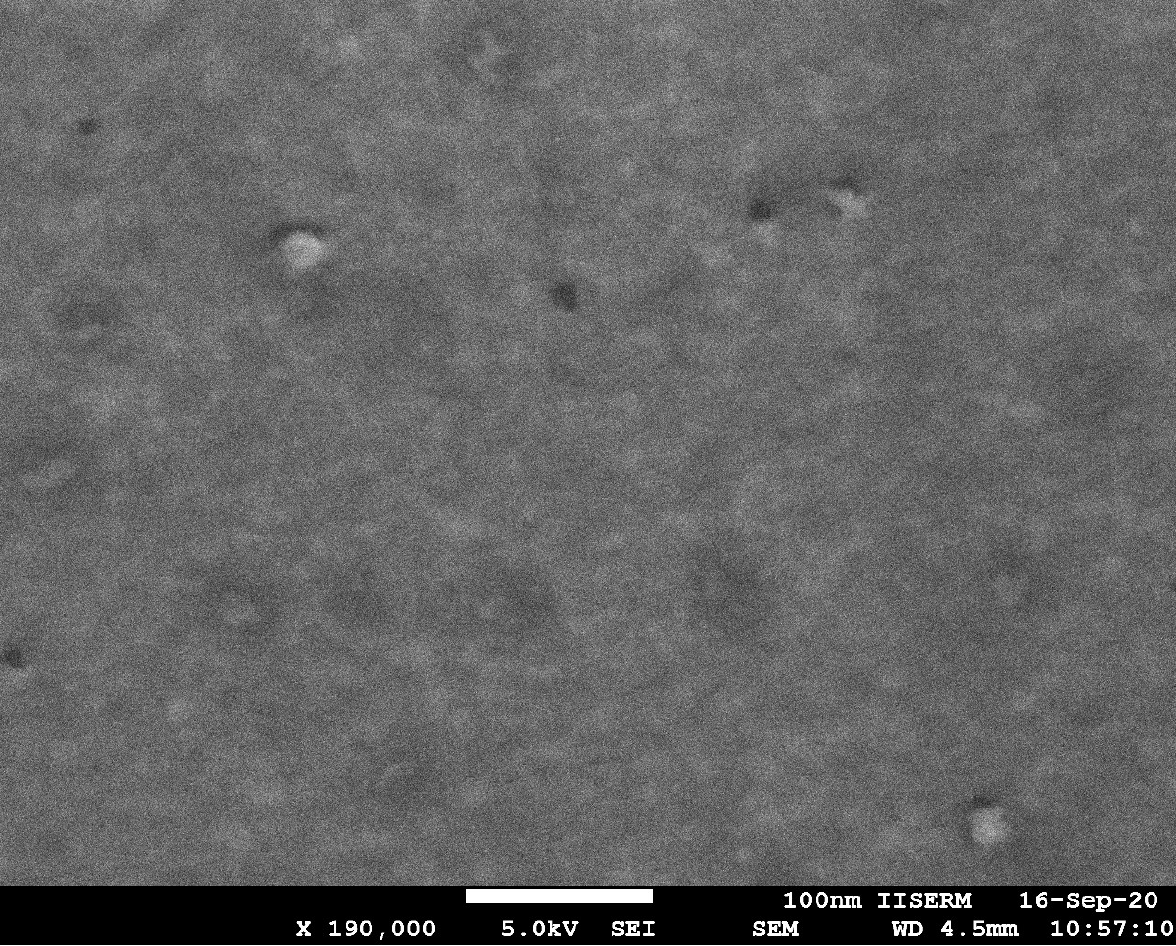}
			\put (5,68) {\large (B)}
			
			\end{overpic}
\end{minipage}

	\caption{ (A) Scanning electron micrograph of grains in $Pd$ films forming the device are shown. (B)Gold has similar sized grains but with less oriented domains and hence a poor contrast in the image.}\label{grain}
 \end{figure}
As cited in the main text reference\cite{resistor} uses the enhanced grain size in $Pd$ films as opposed to other metals like gold  to make shunt resistors for devices like SQUIDs. 
As discussed in the main text it was justifiable that the oriented grains make the anisotropies in the Gr\"{u}neisen parameter enhance due to compression.  

\providecommand{\latin}[1]{#1}
\makeatletter
\providecommand{\doi}
{\begingroup\let\do\@makeother\dospecials
	\catcode`\{=1 \catcode`\}=2 \doi@aux}
\providecommand{\doi@aux}[1]{\endgroup\texttt{#1}}
\makeatother
\providecommand*\mcitethebibliography{\thebibliography}
\csname @ifundefined\endcsname{endmcitethebibliography}
{\let\endmcitethebibliography\endthebibliography}{}